# Simplified Numeric Simulation Approach for $CO_{2,g}$-Water Flow and Trapping at Near-Surface Conditions

AbdAllah A. Youssef, Qi Shao and S. K. Matthäi

Peter Cook Centre for CCS Research & Department of Infrastructure Engineering, The University of Melbourne, Parkville, VIC 3010, Australia

Corresponding authors:

1. AbdAllah A. Youssef, abdallah_youssef@ymail.com
2. Qi Shao, shao.q@unimelb.edu.au

**Abstract**

To simulate $CO_{2,g}$-water flow in tank experiments, subject to viscous, gravitational and capillary forces as well as the dissolution of this gas ($CO_{2,aq}$), we constructed a simple pseudo black-oil model. Simple PVT correlations were used for gas density, viscosity, and solubility as based on experimental studies and equations of state from the literature. These solubility calculations assume instantaneous chemical equilibrium. The applicability of the approach is investigated by modeling the FluidFlower tank experiment (Nordbotten et al., 2022). The simulation captures the expected physical phenomena, including capillary filtration, gravitational segregation, and dissolution fingering. An error in the total mass, due to ignoring solubility variations with pressure remains acceptable as long as the pressure variation in the tank is small.

# 1 Introduction

The continuous increase of anthropogenic greenhouse gases concentration in the atmosphere is classified as the main cause of the global warming (Metz et al., 2005). The highest concentration among these gases is $CO_2$. To meet Paris Climate Agreement adopted in 2015, several approaches have been proposed to store $CO_2$. One of the promising techniques is $CO_2$ geo-sequestration where $CO_2$ is injected into deep saline aquifers or depleted hydrocarbon reservoirs. For better understanding of different physical/chemical processes associated with $CO_2$ flow in porous media, laboratory experiments are conducted.

Fundamentally, $CO_2$ geo-sequestration is a complex process that involves many interacting phenomena at both long- and short-time scales. The latter is more convenient to the laboratory experiments that usually last from several hours to a few days. At this time interval and considering $CO_2$ geo-sequestration in decimeter scale tank, the common processes detected at the laboratory are:

1. gravitational segregation due to large density difference between $CO_2$ and water/brine (e.g., Gilmore et al., 2022);
2. capillary trapping that results from capillary barrier formed at interfaces between facies with different entry pressures where material with higher entry pressure works as an obstacle for non-wetting phase migration (e.g., Ni and Meckel, 2021);
3. mutual solubility of $CO_2$ and water/brine resulting in connective mixing (e.g., Huppert and Neufeld, 2014) and salt precipitation (e.g., Yamamoto and Doughty, 2011).

To formulate a ground truth, the experiments are compared to numerical simulations. The simulation of these simultaneous multi-physics processes usually requires performing compositional simulation. Despite its complex design founded for the sake of predicting interacting processes especially mass transfer between phases, the main disadvantage of compositional simulation is the excessive computational cost. On the contrary, the black-oil simulation is computationally cheap, but less accurate due to ignorance of mutual solubility. Another efficient alternative is the pseudo black-oil simulation.

Irrespective of modeling the associated geo-mechanical events, the simulation of $CO_2$ injection possesses a challenge due to the $CO_2$ compressibility and dissolution. The latter is one of the four trapping mechanisms (Doughty and Pruess, 2004). At reservoir conditions, the dissolution is relatively slow, and its significance is prolonged at the intermediate and long-time scales (Metz et al., 2005). Compared to the reservoir conditions, the dissolution at surface conditions is faster and contributes more to trapping at short time scales. Hence, it is expected that the overall model performance, especially the convective mixing, is very sensitive to the way the solubility is tackled.

The objective of this short note is to describe a simple and computationally efficient pseudo black-oil model, as an alternative to the complex and costly compositional simulation. The presented pseudo black-oil model originates from a basic incompressible-immiscible multi-phase flow model. This approach is convenient for modeling $CO_2$-water tank flow experiments conducted with pressure and temperature ranges close to the standard conditions. This paper proceeds by stating the assumptions that facilitate the simplification of conservation equations.

Then, the gaseous $CO_2$ PVT parameters and solubility are fitted to simple correlations. After that, the proposed approach is adopted to predict the performance of $CO_2$ injection in the FluidFlower rig (Nordbotten et al., 2022). Finally, the limitations and advantages of the presented approach are discussed.

## 2 Methodology
### 2.1 Modeling

Ideally, the $CO_2$-water system is described by a two-component ($CO_2$ and water), two-phase (gas; $g$ and liquid; $l$) system. Considering the absolute pressure and temperature ranges near the surface conditions and their changes across the domain of interest are minute, the following assumptions hold

1. an isothermal system where the changes in the temperature due to Joule-Thomson effect and dissolution quickly calibrate with the surrounding temperature,
2. the change in the formation pore volume is insignificant and the porosity can be assumed invariable unless the compaction is significant,
3. the pure $CO_2$ exists in gas phase while pure water exists as a liquid phase,
4. the densities of the pure component are constant and can be fixed to average values at prescribed datum at the start of the experiment,
5. the vaporized water percentage into gas phase is limited and does not affect properties of gas phase as the water vapor pressure is negligible compared to the $CO_2$,
6. gaseous $CO_2$ dissolves in water forming acidic aqueous solution which has a higher density than pure water,
7. $CO_2$ dissolution in water is limited by the solubility limit which is a function of the pressure and temperature, and finally
8. the changes in viscosities, wettability and surface tension due to the solubility can be neglected.

Based on these assumptions, the system is simplified to a two-phase two-pseudo-component system. The two phases are as previous while the components become $CO_2$ and aqueous. The rich gaseous $CO_2$ phase contains only $CO_2$ component, and the liquid phase is a pseudo component that represents both water and dissolved $CO_2$. Mathematically, the conservation of mass of this system is formulated as two conservation equations with a mass transfer term between two phases, as follows

$$\begin{cases} \phi \rho_g \frac{\partial s_g}{\partial t} = \nabla \cdot \left( \rho_g \lambda_g k \nabla (p_g - \rho_g g D) \right) + (\rho_g q_g)_{s/s} - \Psi \\ \phi \frac{\partial (\rho_l s_l)}{\partial t} = \nabla \cdot \left( \rho_l \lambda_l k \nabla (p_l - \rho_l g D) \right) + \Psi \end{cases} \quad \text{in } (t>0) \cap \Omega \qquad (1),$$

where $\phi$ is porosity, $\rho$ is density, $s$ is saturation, $\lambda$ is the mobility ($\lambda = \frac{k_r}{\mu}$), where $k_r$ is relative permeability and $\mu$ is viscosity, $g$ is the gravitational acceleration, $D$ is depth, and $t$ is time, $\Omega$ is the computational space domain and $\Psi$ is the mass transfer rate due to dissolution of gaseous $CO_2$ into aqueous phase. To complete the description of the system, we need to track the concentration of dissolved $CO_2$ into the liquid phase by solving an advection-diffusion equation. For instance, if the diffusion is neglected that advection equation is stated as follows

$$\phi \frac{\partial (cs_l)}{\partial t} = \nabla \cdot (c\mathbf{v}_l) \qquad (2),$$

where $c = \dfrac{m_{dissolved}}{Vol_l}$ is the concentration of dissolved $CO_2$ into the liquid phase.

## 2.2 Numerical scheme and discretization

The three primary variables that we are after are pressure, saturation and concentration. The system of equations is solved for the three main unknowns with the help of two closure equations: $s_l + s_g = 1$ and $p_c = p_g - p_l$. We use CSMP++ (Complex System Modelling Platform; Matthäi et al., 2007) that is based on IMPES scheme following a sequential algorithm such that the pressure equation is solved implicitly first and then the transport followed by the advection equation are solved explicitly. Before introducing the IMPES form, we perform additional analysis that helps to accelerate the calculations.

We focus on the left-hand side of the aqueous phase equation. By applying the chain rule, the accumulation term is decomposed as

$$\frac{\partial (\rho_l s_l)}{\partial t} = \rho_l \frac{\partial s_l}{\partial t} + s_l \frac{\partial \rho_l}{\partial t} \qquad (3).$$

It is easy to realize that the first term is much larger than the second term ($\rho_l \dfrac{\partial s_l}{\partial t} \gg s_l \dfrac{\partial \rho_l}{\partial t}$). The main reason is that the relative change of the density, as shown later, is very small ($\dfrac{\partial \rho_l}{\rho_l} \approx 10^{-4} : 10^{-3}$) compared to the relative change in the saturation ($\dfrac{\partial s_l}{s_l} \approx 10^{-2} : 10^{-1}$). Therefore, we can propose that $\dfrac{\partial (\rho_l s_l)}{\partial t} \simeq \rho_l \dfrac{\partial s_l}{\partial t}$ and the conservation system (1) reduces to

$$\begin{cases} \phi \rho_g \dfrac{\partial s_g}{\partial t} = -\nabla \cdot (\rho_g \mathbf{v}_g) + (\rho_g q_g)_{s/s} - \Psi \\ \phi \rho_l \dfrac{\partial s_l}{\partial t} = -\nabla \cdot (\rho_l \mathbf{v}_l) + \Psi \end{cases} \quad \text{in } (t > 0) \bigcap \Omega \qquad (4)$$

which is similar to an incompressible system except for the existence of the dissolution term and variation of the liquid density as a consequence of dissolution. If we neglect these exceptions, we obtain the traditional incompressible pressure equation (Aziz and Settari, 1979)

$$\nabla \cdot \left( k\lambda_T \nabla p - k(\lambda_g \rho_g + \lambda_l \rho_l) \mathbf{g} \right) = (q_g)_{s/s} \qquad (5)$$

that results from the sum of the two conservation equations, where $p$ is the global pressure defined as (Chen et al., 2006)

$$p = p_g - \int_{s_l} \left( f_l \frac{dp_c}{d\zeta} \right)(\zeta)d\zeta \qquad (6).$$

Our claim here is that the dissolution term has a small influence on pressure but a major impact on the saturation. Hence, we are solving the incompressible pressure equation (5) followed by the saturation equation (4) that acknowledges the dissolution.

In the current approach, we are proposing that the dissolution reacts as a source/sink term that only affects the saturation treatment. We assume that the dissolution is fast such that gaseous $CO_2$ dissolves instantaneously into aqueous phase in contact with. However, the dissolution is limited by both the solubility limit and the allowable gaseous $CO_2$. These constrains are implanted by adhering to the flow chart shown in Figure 1.

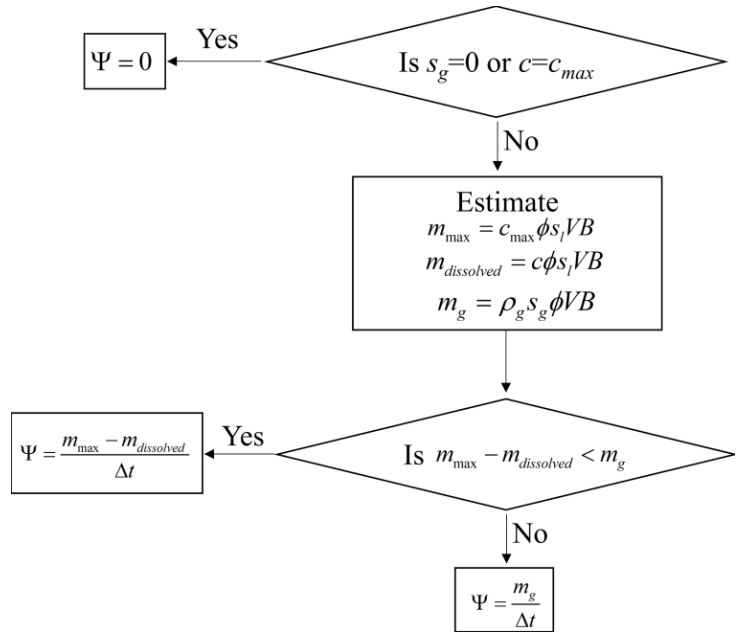

Figure 1. Flow chart of estimation of solubility term $(\Psi)$ in transport equation (4). $m_{max}$ is the maximum allowable dissolved mass corresponding to solubility limit, $m_g$ is the mass of gaseous $CO_2$ and $\Delta t$ is the time step.

In summary, the solution algorithm proceeds as follows

1. define a representative datum and accordingly evaluate pure fluid properties,
2. update pressure to a new time by solving the incompressible pressure equation (5) implicitly based on pure fluid properties evaluated in step 1,
3. calculate $\rho_l$ from the concentration by applying Eq (12); see section 3.2,
4. evaluate $v_l$ from extended multi-phase formula of Darcy's law as $v_l = -k\lambda_l \nabla(p_l - \rho_l gD)$,
5. estimating solubility following the algorithm illustrated in Figure 1,
6. update saturation by solving the liquid transport equation (4) explicitly,

7. update concentration by solving advection equation (2) explicitly, and
8. repeat steps 2-7 till end of simulation time.

## 3 Thermodynamic correlations

Based on the assumption of limited variation of pressure and temperature stated previously in section 2.1, it would be feasible to formulate simple PVT correlations for such narrow ranges. In spite of prescribing constant properties of both pure components, the sensitivity studies and inverse modeling are handier with correlations compared specifying the properties for each run. Herein, we report correlations for $\rho_g, \mu_g$ and $\rho_l$.

### 3.1 Pure gaseous $CO_2$ properties

For pure $CO_2$, two parameters are of interest: viscosity and density. The raw data are generated using Peace software[1]. The interpolation is based on the $CO_2$ properties database reported in Schmidt (2010) that is built utilizing Span and Wagner EOS (Span and Wagner, 1996) and Vesovic et al. (1990) method for transport properties. The parameters are estimated for a pressure range between 101 kPa and 123 kPa and a temperature range between 18 °c and 25 °c (see supplementary material). For the sake of building confidence in the collected raw data before deriving correlations, the data are compared with other data found online and in the literature. The density is assessed by self-comparison with density evaluated using MegaWatSoft calculator[2] that is also based on Span and Wagner EOS (Span and Wagner, 2003). On the other hand, the accuracy of the viscosity raw data is addressed through comparison with two different methods: firstly by self-comparison with the online Engineering Toolbox calculator[3], and secondly by comparison to viscosity evaluated using LMNO calculator[4] that uses Sutherland's formula (Sutherland, 1893).

Figure 2 shows the excellent match of density between Peace software and MegaWatSoft calculator which raises confidence in the collected data. Figure 3 indicates that the $CO_2$ viscosity evaluated from Peace software is perfectly matched with the one estimated from Engineering Toolbox but less than the corresponding viscosity obtained from LMNO calculator by ~$10^{-4}$ cp. This demonstrates that the collected PVT data from Peace software are reliable within the specified pressure and temperatures ranges.

---

[1] Online linear interpolation calculator of $CO_2$ thermodynamic properties released in 2007; http://www.peacesoftware.de/einigewerte/co2_e.html
[2] Online calculator of CO2 properties; https://www.carbon-dioxide-properties.com/default.aspx
[3] https://www.engineeringtoolbox.com/
[4] Online calculator of different fluid properties; https://www.lmnoeng.com/Flow/GasViscosity.php

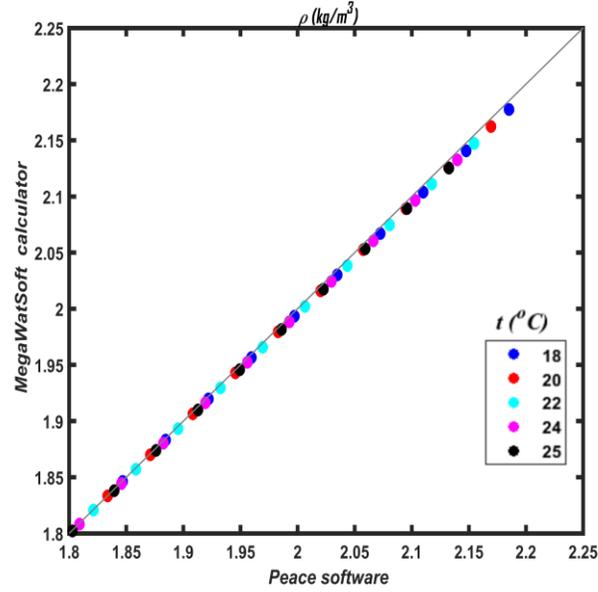

*Figure 2. $CO_2$ density cross plot between Peace software and MegaWatSoft calculator at different temperatures (markers). The solid line is the unit slope line.*

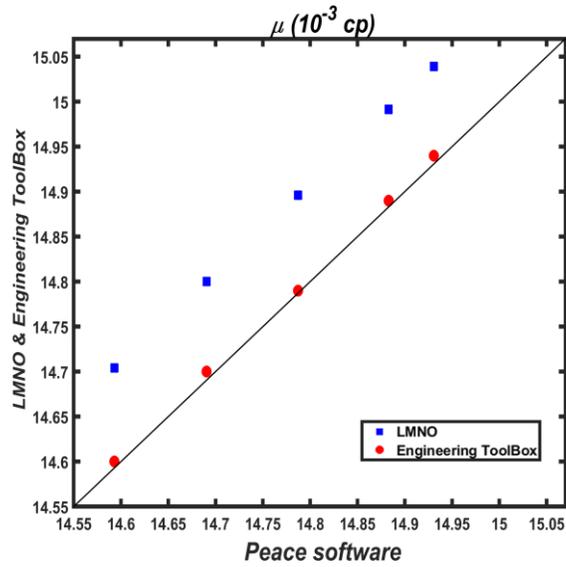

*Figure 3. $CO_2$ viscosity cross plot between Peace software, and LMNO and Engineering ToolBox at different temperatures (markers). The solid line is the unit slope line.*

In order to facilitate the fitting operation, we start by formulating an idea about the behaviour of the variation of $\rho_g$ and $\mu_g$ with the pressure and temperature. Figure 4 confirms the independency of $\mu_g$ on pressure within the range of interest (cf., Sutherland, 1893). Moreover, it has been found that $\mu_g$ varies with the temperature following a third-degree polynomial as follows

$$\mu_g(10^{-3}cp) = a_1 T^2 + a_2 T + a_3 \qquad (7),$$

where $T$ is temperature in °c. In contrast, Figure 4 shows that the density increases linearly with pressure as expected near the standard conditions. Therefore, $\rho_g$ is fitted with the pressure through the straight-line relation

$$\rho_g (\text{kg/m}^3) = ap(\text{kpa}) + b \qquad (8),$$

where $a$ and $b$ are temperature dependent coefficients. It has been found that, within the range of interest, $a$ and $b$ follow a third and fourth degree polynomial variation with the temperature, respectively, as follows

$$\begin{cases} a(\text{kg/m}^3/\text{kpa}) = a_1 T^3 + a_2 T^2 + a_3 T + a_4 \\ b(\text{kg/m}^3) = b_1 T^3 + b_2 T^2 + b_3 T + b_4 \end{cases} \qquad (9).$$

Table 1 summarizes the values of the coefficients of different equations applied in this study.

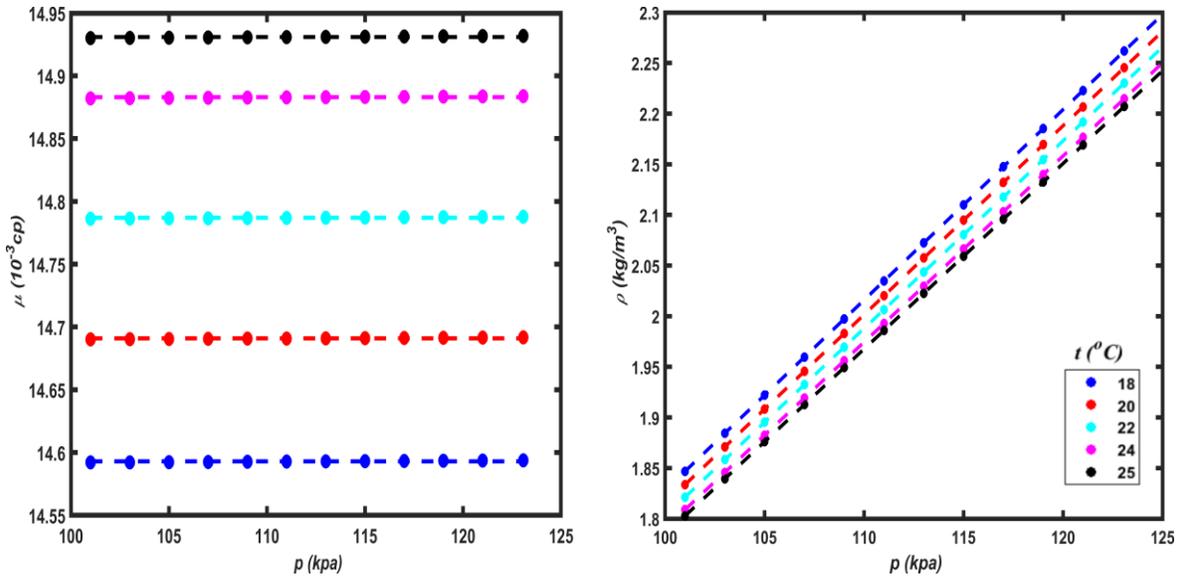

*Figure 4. Variation of pure gas $CO_2$ viscosity (left) and density (right) with pressure at different values of temperature. The markers represent collected data using Peace software and the dashed lines are the best fit.*

*Table 1. Fitting parameters for thermodynamic correlations.*

| Equation | (7) | (9) | (11) |
|---|---|---|---|
| $a_1$ | $-9.5559080095 \times 10^{-5}$ | $-1.75033984146926 \times 10^{-7}$ | $4.0 \times 10^{-7}$ |
| $a_2$ | 0.052347739889 | $1.18625752520443 \times 10^{-5}$ | $-3.8 \times 10^{-5}$ |
| $a_3$ | 13.6817600977 | $-0.000334784869609674$ | 0.00134 |
| $a_4$ |  | 0.0220038529481843 |  |
| $b_1$ |  | $6.30121132429231 \times 10^{-6}$ | $-1.8 \times 10^{-7}$ |
| $b_2$ |  | $-0.000427051888681927$ | $3.9 \times 10^{-6}$ |
| $b_3$ |  | 0.0102882369939649 | $1.2 \times 10^{-4}$ |
| $b_4$ | | -0.135618571701724 | 0.9876 |

## 3.2 Solubility

We propose that the $CO_2$ instantaneously dissolves into the liquid water and forms acidic liquid phase saturated with dissolved $CO_2$, i.e., the liquid phase directly reaches its solubility limit instantaneously as long as there is enough gaseous $CO_2$. In addition, it is assumed that the water only exists in the aqueous phase and does not vaporize into the gaseous $CO_2$ phase which leads to no differentiation between total gaseous pressure and $CO_2$ partial pressure. This approach shows a slight deviation in the estimated miscibility. The estimated error in the worst condition does not exceed 3.2 % within the pressure and temperature limits (Carroll et al., 1991). To minimize this error, the solubility is correlated to the partial pressure instead the total pressure. Based on the filtration of the experimental data conducted by Carroll et al. (1991), we propose a power trend between the solubility and the pressure as follows

$$sol(\text{mol}\%) = ap^b \qquad (10),$$

again $a$ and $b$ are temperature dependent parameters fitted versus quadratic and cubic polynomials, respectively, as follows

$$\begin{cases} a(\text{kPa}^{-1}) = a_1 T^2 + a_2 T + a_3 \\ b = b_1 T^3 + b_2 T^2 + b_3 T + b_4 \end{cases} \qquad (11).$$

Figure 5 indicates that the proposed trend fits the solubility estimated by Carroll et al. (1991) with a resendable accuracy.

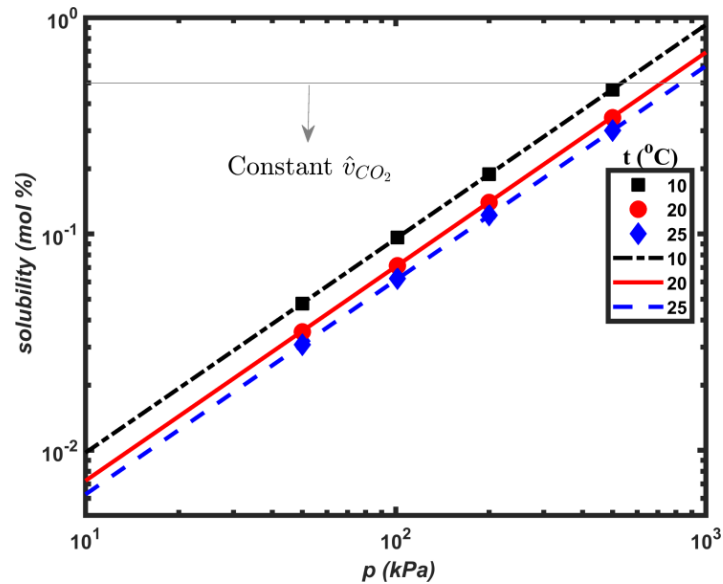

*Figure 5. Best-fit power relation based on Eq (10) between pressure and $CO_2$ solubility in water at different temperatures (dashed lines). The markers represent the data estimated based on the approach proposed by Carroll et al. (1991).*

The dissolution of $CO_2$ into liquid phase increases the liquid mass in addition to its volume. Overall, the increase in the mass is larger than the corresponding increase in the volume. Therefore,

$\rho_l$ increases by the solubility of $CO_2$. To derive a correlation between $\rho_l$ and $c$, it is necessary to know how the solubility affects the partial molar volume and if such effect is significant or minute. For a solubility less than 0.5 mol%, Parkinson and De Nevers (1969) found that the partial molar volume of the dissolved $CO_2$ ($v_{CO_2}$) is independent on temperature and is equivalent to $0.0376 \frac{m^3}{kg-mol}$. Within the previously estimated pressure and temperature ranges, the solubility limit is less than 0.5 mol% (Figure 5). Neglecting the change in the partial molar volume of water, the density of the liquid phase can be written as (Appendix)

$$\rho_l = \rho_{H_2O} + (1 - \frac{v_{CO_2}\rho_{H_2O}}{M_{CO_2}})c \qquad (12),$$

where $M$ is the molecule weight and $\rho_{H_2O}$ is the density of the pure water. At the solubility limit, the maximum concentration is evaluated from solubility as

$$c_{max} = \frac{\rho_{H_2O}M_{CO_2}sol(mol\%)}{v_{CO_2}\rho_{H_2O}sol(mol\%) + [100 - sol(mol\%)]M_{H_2O}} \qquad (13).$$

Figure 6 shows the variation of $c_{max}$ with the pressure in the range where $v_{CO_2}$ is constant.

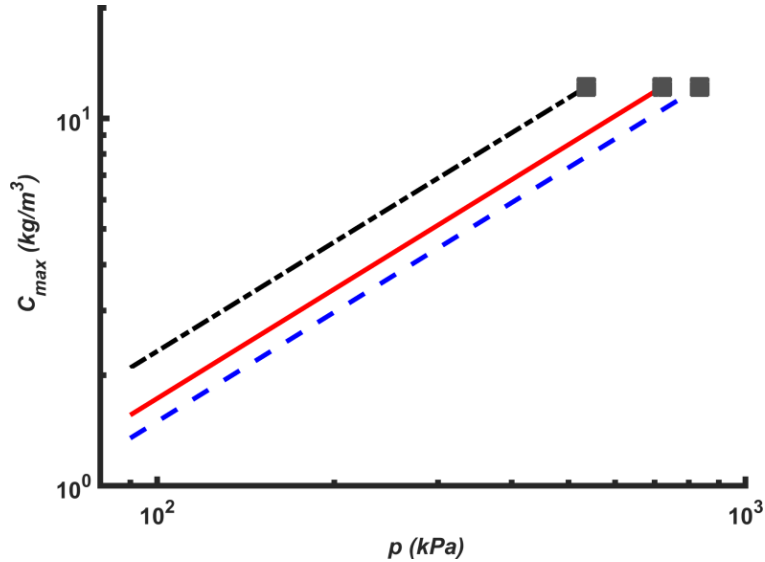

*Figure 6. Change of aqueous phase maximum concentration with pressure considering that the water density is invariant with pressure and constant at 1002 kg/m³. The markers show the limit of the relation where $v_{CO_2}$ is not anymore constant.*

## 4 Results

The previously described modeling choices have been coded into the Complex System Modelling Platform (CSMP++; Matthäi et al., 2007; Paluszny et al., 2007) discretizing the governing equations on a collocated finite element – finite volume (FE-FV) mesh using operator splitting and a sequential solution approach. The code was then used to simulate flow in FluidFlower rig following Nordbotten et al. (2022) proposal. The petrophysical parameters for each sand are summarized in Table 2. Figure 7 shows the status of both free and dissolved $CO_2$ after 24 hours of simulation. From a qualitative perspective, the results indicate the capability of the proposed modeling choices for simulating the gravity segregation, capillary trapping and convective dissolution in a reasonable way. The comparison between the simulation performance and the real experiment (Figure 8) reveals two important features:

1. the specified capillary pressure in the simulation is larger than the actual values, hence, free $CO_2$ is delayed from migrating upward in the top section of the rig,
2. the dissolution is slower compared to the actual dissolution rate and the advection is slower.

*Table 2. Petrophysical parameters of different sands used in FluidFlower benchmark study.*

| Sand name | Permeability (m$^2$) | Porosity |
|---|---|---|
| ESF | $7.40\times10^{-11}$ | 0.43 |
| C | $3.45\times10^{-10}$ | 0.44 |
| D | $4.18\times10^{-10}$ | 0.44 |
| E | $7.40\times10^{-10}$ | 0.45 |
| F | $1.09\times10^{-09}$ | 0.45 |
| G | $1.18\times10^{-09}$ | 0.44 |

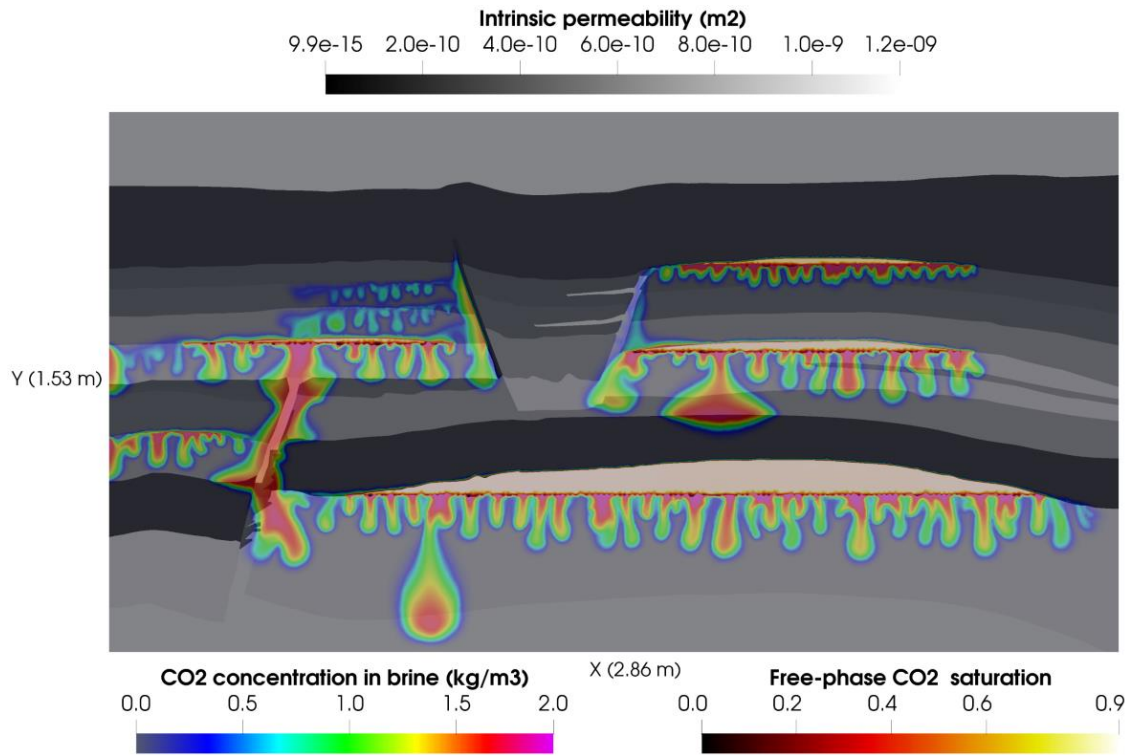

*Figure 7. Simulated free $CO_2$ saturation and dissolved $CO_2$ distributions after 1 day from start of injection in FluidFlower tank experiment with absolute permeability shown in the background.*

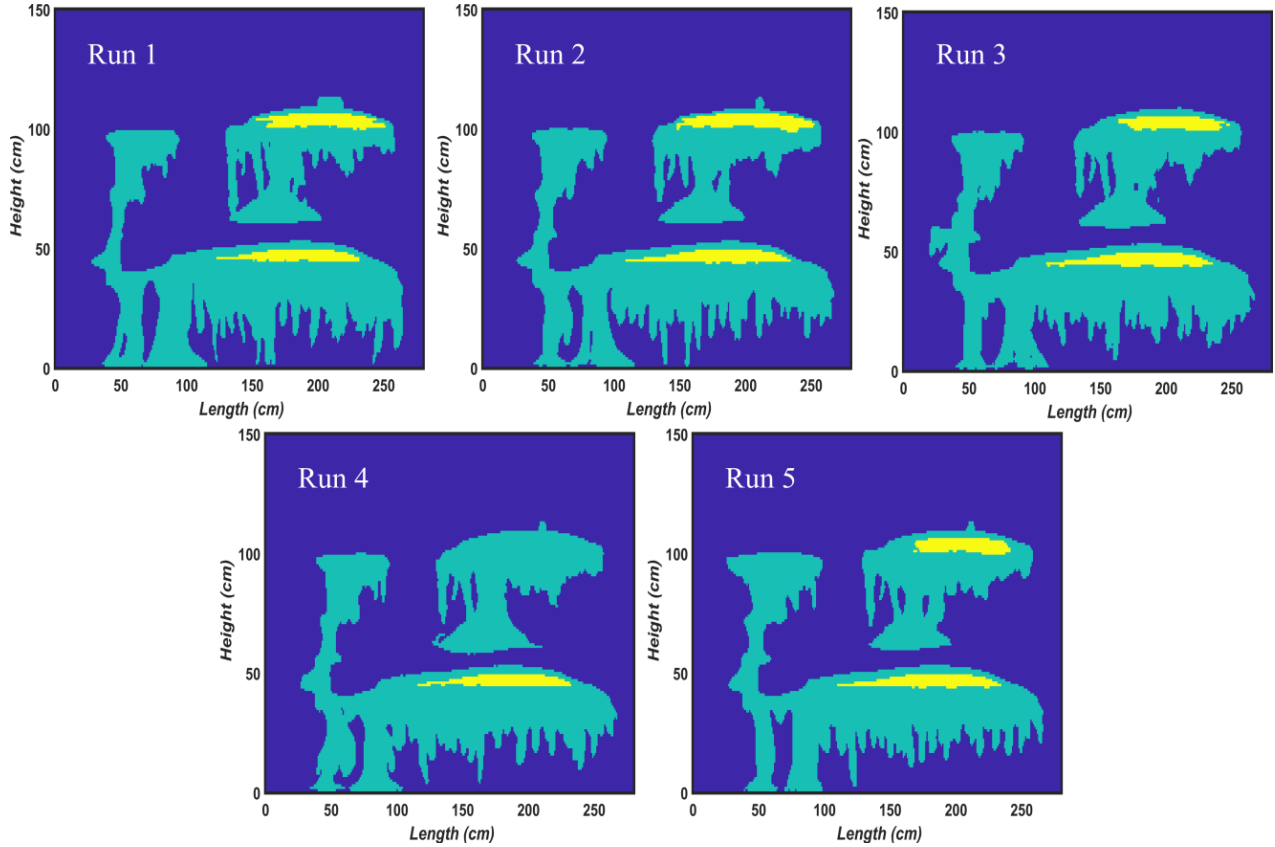

*Figure 8. Segmentation distribution after 1 day of five out of six experiments conducted in FluidFlower rig following benchmark regulations* (Nordbotten et al., 2022). *The yellow color represrns free $CO_2$, the dark blue color shows fresh water, and the light blue is for water that contains dissolved $CO_2$.*

## 5 Discussion

The approach that we follow in our simplification leads to a system similar to the incompressible system except for the inclusion of the dissolution term. Theoretically, this system is very restrictive to a balance between the mass and the volume, and the thermodynamic relation between the concentration and density [Eq (12)]. Therefore, it is expected that fixing the gas density violates the mass balance. The question that comes to the light is how far we are from the correct results. To give an answer to this question, we compare the mass of $CO_2$ in the system with the injected mass. Figure 9 indicates that the current analysis overestimates the mass of $CO_2$ by less than 10%, except at the spike at the activation of the second port, compared to the expected mass. Hence, the approximation suggested is acceptable and the error is controlled as far as the experiment only lasts for a relatively short time (e.g., a few days).

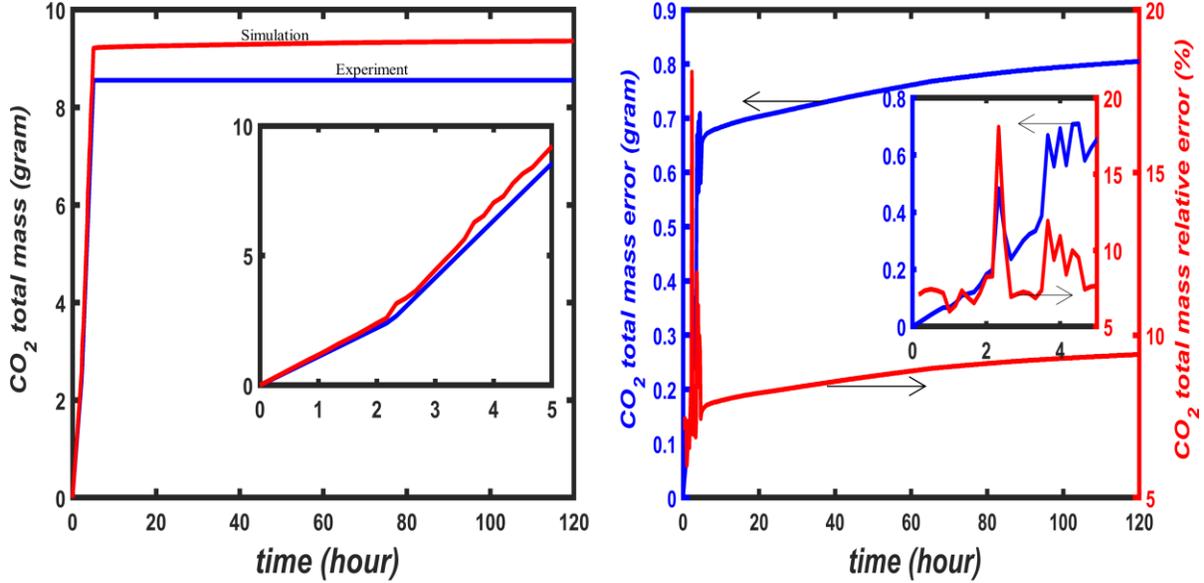

*Figure 9. Evolution of CO₂ total mass during FluidFlower tank experiment (left), and corresponding CO₂ total mass absolute error and relative error (right). Docked figures show a zoom to injections interval.*

One of the achievements of this study is providing a quick estimation of $CO_2$ dissolution from the gas phase to liquid phase. The approach presented here is based on two main assumptions. The first one is that dissolution is instantaneous such that the dissolution rate can be considered infinite. This assumption should not be taken as granted. The change of the conditions from surface to reservoir shall be considered as they play a significant role in the determination of dissolution rate. The second one is that the algorithm presented in Figure 1 considers only change in concentration based on the change of mass, while the change in the volume is treated across the solution of the transport equation. A more rigorous way would be to evaluate it through an iterative process with the transport equation, however, this comes with increased computational cost and time.

## 6 Conclusion

This study presents a non-sophisticated approach for modelling $CO_2$ injection in tank experiments where the operating conditions are close to the surface conditions. The approach relies on solving an incompressible pressure equation and treating the dissolution term as a source term into the transport equation. The approach has the advantage of low computational cost as compared to the compositional model.

## Appendix

This appendix shows the derivation of Eq (12). The density of the liquid phase can be formulated as

$$\rho_l = \frac{m_{CO_2} + m_w}{vol_l} = c + \frac{m_w}{vol_l} \qquad (A1).$$

Considering the relations between mass and number of moles, and the relation between volume and partial molar volumes, Eq (A1) can be rewritten as follows

$$\rho_l = c + \frac{M_w}{\frac{n_{CO_2} v_{CO_2}}{n_w} + v_w} \tag{A2},$$

where $n$ is the number of moles. The components mole ratio can be easily formulated as

$$\frac{n_{CO_2}}{n_w} = \frac{m_{CO_2} M_w}{m_w M_{CO_2}} = \frac{cM_w}{M_{CO_2}(m_w/vol_l)} \tag{A3}.$$

Substituting by (A1), Eq (A3) can be rewritten as

$$\frac{n_{CO_2}}{n_w} = \frac{cM_w}{M_{CO_2}(\rho_l - c)} \tag{A4}.$$

Substituting (A4) into (A2) and manipulating, the linear function (12) of $\rho_l$ in $c$ can be derived.